\documentclass{PoS}

\usepackage{amsmath,amsfonts,amssymb}

\usepackage{slashed}

\def\one{\mbox{1 \kern-.59em {\rm l}}}

 \def\N{{\rm I\mkern-3mu N}}
\def\R{{\rm I\mkern-3mu R}}
\def\Q{{\@QC Q}}
\def\C{{\@QC C}}
\def\@QC#1{\mathpalette{\setbox0=\hbox\bgroup$\rm}%
  {\egroup C$\egroup\rm\rlap{\kern0.4\wd0\vrule
  width 0.05\wd0 height 0.97\ht0 depth -0.01\ht0}%
  #1\bgroup}}

\def\cN{{\cal N}}
\def\cM{{\cal M}}

\def\cC{{\cal C}}
\def\cJ{{\cal J}}

\newcommand{\msu}{\mathfrak{s}\mathfrak{u}}

\def\nn{\nonumber}
\def\bea{\begin{eqnarray}}
\def\eea{\end{eqnarray}}
\def\be{\begin{equation}}
\def\ee{\end{equation}}

\newcommand{\eq}[1]{(\ref{#1})}

\def\a{\alpha}          

\def\d{\delta}

 \def\L{\Lambda} 

\renewcommand{\L}{\Lambda}

\def\cA{{\cal A}}  \def\cC{{\cal C}}
  
 \def\cH{{\cal H}} \def\cI{{\cal I}}
\def\cJ{{\cal J}}  \def\cL{{\cal L}}
\def\cM{{\cal M}} \def\cN{{\cal N}}

\def\LNC{\L_{\rm{NC}}}
\def\Mat{\rm{Mat}}
\def\del{\partial}
\def\bDelta{{\bf\Delta}}
 \def\bDelta{{\Box}}

\title{On matrix geometry}

\ShortTitle{On matrix geometry}

\author{\speaker{Harold Steinacker}\\
        Fakult\"at f\"ur Physik, Universit\"at Wien \\
Boltzmanngasse 5, A-1090 Wien, Austria\\
        E-mail: \email{harold.steinacker@univie.ac.at}}

\abstract{The foundations of matrix geometry are discussed, which provides the basis for 
recent progress on the effective geometry and gravity in Yang-Mills matrix models.
Basic examples lead to a notion of embedded noncommutative spaces (branes)
with emergent Riemannian geometry. This class of configurations turns out to be
preserved under small deformations, and is therefore
appropriate for matrix models. The relation with spectral geometry is discussed.
A possible realization of sufficiently generic 4-dimensional geometries as noncommutative branes 
in $D=10$ matrix models is sketched.}

\FullConference{Corfu Summer Institute on Elementary Particles and Physics - Workshop on Non 			Commutative Field Theory and Gravity,\\
		September 8-12, 2010\\
		Corfu Greece}

\begin{document}

\section{Introduction}

One of the most fascinating ideas in recent years is the proposal 
that matrix models of Yang-Mills type, in particular 
certain models which have been put forward in string theory \cite{Ishibashi:1996xs,Banks:1996vh}, may provide a description 
for the quantum structure of space-time and geometry. 
The beauty of the proposal lies in the 
simplicity of these models, whose structure  is essentially
\be
S = {\rm Tr} [X^a,X^b][X^{a'},X^{b'}] g_{aa'} g_{bb'} \,\, + {\rm fermions} .
\ee
Here $X^a,\, a= 1,...,D$ are a set of hermitian matrices, and we restrict ourselves to the case of Euclidean signature 
with $g_{ab} = \d_{ab}$  in this article. No notion of differential geometry 
and classical space-time whatsoever is used in this action. 
The geometrical structures arise in a certain ``semi-classical limit'',
in terms of solutions of these models. 
The aim of this article is to clarify the scope and the mathematical description of this ``emergent`` geometry.

Simple examples of such matrix geometries, notably the fuzzy sphere $S^2_N$ or more general quantized homogeneous
spaces including the Moyal-Weyl quantum plane $\R^{2n}_\theta$, have been studied in great detail.
However in order to describe the general geometries required for gravity, one cannot rely on the special group-theoretical 
structure of these simple examples. This obstacle was removed recently by realizing
\cite{Steinacker:2010rh,Steinacker:2008ri} that there is a sufficiently large class of
matrix geometries with generic geometry, which can play the role of space-time (possibly with extra dimensions)
in Yang-Mills matrix models. 
The key to understand their geometry is to realize that these generic matrix geometries should be considered
as {\em embedded noncommutative (NC) spaces} resp. NC branes
$\cM \subset \R^D$. The effective geometry is very clear in the ''semi-classical limit`` of the matrix geometry, 
where commutators are replaced by Poisson brackets. As we will recall below,
$\cM$ then inherits the pull-back metric $g_{\mu\nu}$ of $\R^D$, which combines 
with the Poisson (or symplectic) structure $\theta^{\mu\nu}(x)$ to form an effective metric $G_{\mu\nu}(x)$, much like
the open string metric in string theory. 

However, the description of matrix geometries in terms of a semi-classical limit is based on certain assumptions,
and is may seem desirable
to have a more precise characterization of ``matrix geometries``. This is possible in the 
simple examples mentioned above, introducing e.g. a differential calculus and using ideas of 
noncommutative differential geometry. The standard realization of these structures
for the homogeneous spaces \cite{Madore:1991bw,Grosse:2000gd,Balachandran:2005ew} however relies on the
special group-theoretical structure. Moreover, it turns out that this differential calculus is essentially that of
the ambient space $\R^D$, and does not know about the intrinsic geometry of the fuzzy space. 
One might consider to use Connes NC differential calculus, 
which however is not naturally adapted to fuzzy geometries (e.g. there is an issue with chirality).
Instead, we will follow an important lesson from Yang-Mills matrix models:
there is no need for 
any additional structure, the models contain all the ingredients required for physics. Our task is merely to 
extract this information without mathematical prejudice. This is the strategy adopted here.

In this article, various aspects of matrix geometry arising in Yang-Mills matrix models 
and their mathematical description will be discussed.
We first recall some examples of finite matrix geometries, described by 
finite-dimensional matrix algebras $\cA = \Mat(N,\C)$. This includes a series of very clear and well-known 
examples such as the fuzzy sphere $S^2_N$, which can be considered for any $N \in \N$. However matrix geometries are much more 
general, and also cover singular geometries such as intersecting branes. Transitions between 
different topologies are conceivable, and physically very interesting.
Therefore the meaning of matrix geometry and even topology can in general only be approximate, since for small $N$ there can 
be no way to exactly separate and characterize them. 
The appropriate concept is that of an effective or ''emergent`` geometry, which is valid
within a certain range of energies. 
This is entirely sufficient for any physical application and in fact to be expected, since the Planck scale 
provides a natural limitation for geometry. Our task is therefore to 
find an appropriate and useful description of generic matrix geometries, where mathematical axioms on the 
geometry are replaced by estimates on the validity of certain effective descriptions.
We will discuss the appropriate tools for this description here, delegating the estimates for future work.

\section{Examples of matrix geometries}

\subsection{Prototype: the fuzzy sphere}

The fuzzy sphere  $S^2_N$  \cite{Madore:1991bw,hoppe} is a 
quantization resp. matrix approximation of the usual sphere $S^2$,
with a cutoff in the angular momentum.
We first note that the algebra of functions on the ordinary
sphere can be generated by the coordinate functions $x^a$ of $\R^3$ modulo the
relation $ \sum_{ {a}=1}^{3} {x}^{a}{x}^{a} = 1$. 
The fuzzy sphere $S^2_{N}$ is a
non-commutative space defined in terms of three $N \times N$ hermitian matrices $X^a, a=1,2,3$ 
subject to the relations
\begin{equation}
[ X^{{a}}, X^{{b}} ] = \frac{i}{\sqrt{C_N}}\varepsilon^{abc}\, X^{{c}}~ , 
\qquad \sum_{{a}=1}^{3} X^{{a}} X^{{a}} =  \one 
\end{equation}
where $C_N= \frac 14(N^2-1)$ is the value of the quadratic Casimir of $\msu(2)$ on $\C^N$.
They are realized by the generators of the $N$-dimensional representation $(N)$ of $\msu(2)$. 
The matrices $X^a$ should be interpreted as quantized embedding functions
in the Euclidean space $\R^3$,
\be
X^a \sim x^a:\quad S^2 \hookrightarrow \R^3.
\ee
They generate an algebra $\cA \cong \Mat(N,\C)$,
which should be viewed as quantized algebra of functions on the symplectic space $(S^2,\omega_N)$
where $\omega_N$ is the canonical $SU(2)$-invariant symplectic form on $S^2$
with $\int \omega_N = 2\pi N$. 
The best way to see this is to decompose $\cA$ into irreps under the adjoint action of $SU(2)$, 
which is obtained from 
\bea
S^2_{N} \cong (N) \otimes (\bar N) 
&=& (1) \oplus (3) \oplus ... \oplus (2N-1) \nn\\
&=& \{\hat Y^{0}_0\} \,\oplus \, ... \, \oplus\, \{\hat Y^{N-1}_m\}.
\label{fuzzyharmonics}
\eea
This provides the definition of the 
fuzzy spherical harmonics $\hat Y^{l}_m$, and defines the {\em quantization map}
\be
\begin{array}{rcl}
\cI: \quad \cC(S^2) &\to& \cA\,\,= \,\, \Mat(N,\C)\, \\
 Y^l_m &\mapsto& \left\{\begin{array}{c}
                         \hat Y^l_m, \quad l<N \\ 0, \quad l \geq N
                        \end{array}\right.

\end{array}
\label{quant-map-S^2} 
\ee
It follows easily that $\cI(i\{x^a,x^b\}) = [X^a,X^b]$ where $\{,\}$ denotes the Poisson brackets 
corresponding to the symplectic form $\omega_N = \frac N2 \varepsilon_{abc} x^a dx^b dx^c$
on $S^2$.
Together with the fact that $\cI(f g) \to \cI(f)\cI(g)$ for $N \to \infty$ (which is not hard to prove), 
$\cI(i\{f,g\}) \approx [\cI(f),\cI(g)]$ follows. This means that $S^2_N$  is the quantization of $(S^2,\omega_N)$.

Moreover, there is a natural Laplace operator on $S^2_N$ defined as
\be
{\bDelta} = [X^a,[X^b,.]]\d_{ab}
\label{matrix-laplacian-S2}
\ee
which is invariant under $SO(3)$.
Its spectrum coincides with the spectrum of the classical 
Laplace operator on $S^2$ up to the cutoff, and the eigenvectors are given by the fuzzy spherical harmonics 
$\hat Y^l_m$.

In this special example, \eq{fuzzyharmonics} allows to construct a 
 a series of embeddings
\be
\cA_N \subset \cA_{N+1} \subset ...   
\label{fuzzy-sequence}
\ee
with norm-preserving embedding maps. 
This allows to recover the classical sphere by taking the inductive limit.
While this is a very nice structure, we do not want to rely on the existence of such 
series of embeddings, for reasons explained below.

\subsection{Other examples}

A straightforward generalization of the fuzzy sphere leads to the {\em fuzzy complex projective space} 
$\C P^{n-1}_N$, which is defined in terms of hermitian matrices $X^a$, $a = 1,2,...,n^2-1$  subject to the 
relations 
\be
[ X^{{a}}, X^{{b}} ] = \frac{i}{\sqrt{C_N'}} f^{ab}_c\, X^{{c}}~, \qquad 
 d_{ab}^c  X^{{a}} X^{{b}} = D_N X^c, \qquad X_a X^a =  \one 
\ee
(adopting a sum convention).
Here $f^{ab}_c$ are the structure constants of $\msu(n)$, $d^{abc}$ is the totally symmetric invariant tensor,
and $C_N', D_N$ are group-theoretical constants which are not needed here.
These relations are realized by the generators of  $\msu(n)$ acting on irreducible representations $\C^{d_N}$
with highest weight $(N,0,...,0)$ or $(0,0,...,N)$. Again, 
the matrices $X^a$ should be interpreted as quantized embedding functions
in the Euclidean space $\msu(n) \cong \R^{n^2-1}$,
\be
X^a \sim x^a: \quad\C P^{n-1} \hookrightarrow \R^{n^2-1}.
\ee
They generate an algebra $\cA \cong \Mat(d_N,\C)$,
which should be viewed as quantized algebra of functions on 
the symplectic space $(\C P^{n-1}, N\omega)$
where $\omega$ is the canonical $SU(n)$-invariant symplectic form on $\C P^{n-1}$.
It is easy to write down a quantization map analogous to \eq{quant-map-S^2}, 
\be
\cI: \quad \cC(\C P^{n-1}) \to \cA\,
\label{quant-map-CP}
\ee
using the decomposition of $\cA$ into irreps of $\msu(n)$. 
Again, there is a natural Laplace operator on $\C P^{n-1}_N$ defined as in \eq{matrix-laplacian-S2}
whose spectrum coincides with the classical one up to the cutoff.
A similar construction can be given for any coadjoint orbit 
of a compact Lie group.

The {\em fuzzy torus} $T^2_\theta$ can be defined  in terms of clock- and shift operators 
$U,V$ acting on $\C^N$ with relations $U V = q V U$
for $q^N=1$, with $U^N = V^N = 1$. 
However one can also view it as embedded noncommutative space in $\R^4$,
by defining  4 hermitian matrices
$U = X^1 + i X^2, \,\, V = X^3 + i X^4$ which satisfy the relations
\bea
(X^1)^2 + (X^2)^2 &=& 1 = (X^3)^2 + (X^4)^2 ,  \nn\\
(X^1 + i X^2) (X^3 + i X^4) &=& q (X^3 + i X^4)(X^1 + i X^2) .
\label{fuzzy-torus}
\eea
They can again be viewed as embedding maps
\be
X^a \sim x^a: \quad T^2 \hookrightarrow \R^{4} .
\ee 
These matrices generate 
the algebra $\cA \cong \Mat(N,\C)$, which using the decomposition into irreps 
under $U(1) \times U(1)$ can be viewed as quantization of the function algebra $\cC(T^2)$
on the symplectic space $(T^2,\omega_N)$.
The spectrum of the matrix Laplacian \eq{matrix-laplacian-S2} approximately 
coincides with the classical case below the cutoff.

Finally, the {\em Moyal-Weyl quantum plane} $\R^{2n}_\theta$
is defined in terms of $2n$ (infinite-dimensional) hermitian matrices $X^a \in \cL(\cH)$
subject to the relations
\be
[X^\mu,X^\nu] = i \theta^{\mu\nu} \one
\ee
where $\theta^{\mu\nu} = - \theta^{\nu\mu} \in \R$. Here $\cH$ is a separable Hilbert space.
This generates the ($n$-dimensional) Heisenberg algebra $\cA$ (or some suitable 
refinement of it, ignoring operator-technical subtleties here), 
which can be viewed as quantization of the algebra of functions on $\R^{2n}$
using e.g. the Weyl quantization map\footnote{which in turn is defined in terms of 
plane waves i.e. irreducible representations of the translation group.}. 
Of course, the matrices $X^\mu$ should be viewed as quantizations of the classical
coordinate functions $X^\mu \sim x^\mu:\,\, \R^{2n} \to \R^{2n}$.
Again the Laplacian \eq{matrix-laplacian-S2} coincides with the classical one,
for the effective metric specified in \eq{G-def-general}.

This concludes our brief exhibition of matrix geometries, whose geometry is obvious
because of their symmetry. We will learn below how to generalize them for generic geometries,
and how to systematically extract their geometry without using this symmetry.
 Besides these and other nice examples, 
there are also more exotic and singular spaces
that can be modeled by matrices, such as intersecting spaces, stacks of spaces, etc.

\subsection{Lessons and cautions}

We draw the following general lessons from the above examples:
\begin{itemize}

\item
The algebra $\cA = \cL(\cH)$ of linear operators on $\cH$
should be viewed as quantization of the algebra of functions 
on some symplectic space $(\cM,\omega)$. 
However as abstract algebra, $\cA$ carries no geometrical information, not even the dimension or the topology of 
the corresponding space.  $\dim(\cH)$ merely counts the number of ``quantum cells'', more precisely it 
measures the volume via the  semi-classical relation (cf. \eq{quant-map-2})
\be
\int\frac{\omega^n}{n!} f \,\sim \,  (2\pi)^n Tr \cI(f)
\ee
\item
The geometrical information is encoded in the {\em specific matrices} $X^a$, which 
should be interpreted as embedding functions 
\be
X^a \, \sim \, x^a: \cM \hookrightarrow \R^D .
\ee
They encode the embedding geometry, which is contained e.g. in the 
matrix Laplacian \eq{matrix-laplacian-S2}. We will learn below how to extract this more 
directly.
The Poisson resp. symplectic structure is encoded in their commutation relations.
In this way,
even finite-dimensional matrices can describe various geometries to a high precision.

\item
In some sense, every non-degenerate and ``regular'' fuzzy space 
locally looks like the quantization of some Poisson manifold, in particular
like some Moyal-Weyl quantum plane $\R^{2n}_\theta$. The algebra of functions on
$\R^{2n}_\theta$ is infinite-dimensional only because its volume is infinite.
For example,
$\C P^n_N$ can be viewed as particular compactification of $\R^{2n}_\theta$.

\end{itemize}

This leads to the idea that generic geometries can be described similarly as 
{\em embedded noncommutative spaces} in matrix models, interpreting the matrices $X^a$ as 
 quantized embedding maps $X^a \sim x^a:\,\,\cM \hookrightarrow \R^D$.
However, some cautionary remarks on matrix geometries are in order.

The problem of identifying the geometry corresponding to some given
configuration $\{X^a\}$ in the matrix model is clearly hard\footnote{A priori one does not have a sequence of matrices
as in \eq{fuzzy-sequence}.}, since general matrices do not necessarily admit a geometrical interpretation.
There is not even a notion of dimension at this level of generality. In fact 
matrix models can describe much more general situations,  
such as multiple submanifolds (''branes``),
intersecting branes, manifolds suspended between
branes, etc., essentially the whole zoo of string theory.
Each of these are very interesting and
should be treated separately. Therefore we have to make some simplifying assumptions, 
and focus on the simplest case of 
classical submanifolds (and possibly stacks of coinciding branes.).
Indeed, there is a large class of configurations 
which clearly have such a geometrical interpretation. 
For example, we will show in section \ref{sec:generic-4D} how to realize 
a large class of generic 4d geometries through such matrix geometries. 

A sharp separation between admissible and non-admissible matrix geometries would in fact be 
inappropriate in the context of matrix model, whose main merit is the definition of 
quantization in terms of an integral over the space of {\em all} matrices, 
\be
Z = \int dX^a e^{-S[X]} 
\label{pathintegral}
\ee
and similarly for correlation functions. The ultimate aim is to show that the dominant
contributions to this integral correspond to matrix configurations which have a 
geometrical meaning and are relevant to physics. However, the integral is over all possible matrices,
including geometries with different dimensions and topologies.
It is therefore clear that such a geometric notion can only be approximate or ``emergent''.

Finally, we want to address the issue of finite-dimensional versus infinite-dimensional 
matrix algebras.
Imagine that our space-time was fuzzy, with an area quantization characterized by 
the scale $\LNC$ (one may expect $\LNC \approx \L_{\rm Planck}$), 
and perhaps even compact of size $R$ (e.g. with the topology of $T^4$). 
Then there would be only finitely many ``quantum cells'', and the geometry should 
be modeled by some finite $N$ -dimensional (matrix) algebra. No experiment on earth,
not even at CERN, can directly access the Planck scale, and all 
measurements about the geometry could be in perfect agreement with such a model
in terms of a finite matrix geometry. Therefore the limit
$N \to \infty$ is not essential for local physics, however there must be a large ``separation of scales``. 
Let   $\L_{\rm cosm} \sim 1/R$ some cosmological scale. 
Then as long as $\L_{\rm cosm} \ll \LNC$ and $\L_{\rm phys} \ll \LNC$ where 
$\L_{\rm phys}$ is the maximal available energy for experiments, 
then a description in terms of finite matrix geometries should be perfectly
adequate. The effective geometry would be arguably the same if the spectrum of the 
corresponding fuzzy Dirac or Laplace operator
approximately coincides with the continuum case {\em up to energies of order $\LNC$}.
There is no obvious requirement above that scale.

\section{Spectral matrix geometry}

We want to understand more generally such ``matrix geometries'', described by a number 
of hermitian matrices $X^a \in \cA= \cL(\cH)$. Here $\cH$ is a finite-dimensional
or infinite-dimensional (separable) Hilbert space.

From an algebraic point of view, such
matrix algebras are quite boring and
in a sense trivial. In fact, Wedderburns theorem implies that the algebra generated by 
finite-dimensional hermitian matrices is 
always the product of simple matrix algebras. However, the point is that 
even simple matrix algebras can describe non-trivial geometries, as demonstrated by the above examples.
It is the additional structure provided by the specific embedding matrices $X^a$ which
makes such matrix geometries interesting and non-trivial.

One way to extract geometrical information from a space $\cM$ which naturally generalizes to
the noncommutative setting is via spectral geometry. 
In the classical case, one can
consider the heat kernel expansion of the Laplacian $\Delta_g$  of a compact Riemannian manifolds $(\cM,g)$
\cite{Gilkey:1995mj},
\be
{\rm Tr} e^{-\a \Delta_g} = \sum_{n\geq0} \a^{(n-d)/2} \int_\cM d^d x \sqrt{|g|}\, a_n(x) .
\ee
The Seeley-de Witt coefficients $a_n(x)$ of this asymptotic expansion are determined by the intrinsic geometry of $\cM$, 
e.g. $a_2 \sim -\frac{R[g]}6$ where $R[g]$ is the curvature scalar. This provides physically valuable 
information on $\cM$, and describes the one-loop effective action.
In particular, the leading term allows to compute the number of eigenvalues below some cutoff,
\be
\cN_\Delta(\L) := \#\{\mu^2\in {\rm spec} \Delta;\, \mu^2\leq \L^2 \} .
\ee
dropping the subscript $g$ of the Laplacian.
One obtains Weyls famous asymptotic formula
\be
\cN_\Delta(\L) \sim c_d{\rm vol}\cM \,\L^{d} , \qquad c_d = \frac{{\rm vol} S^{d-1}}{d(2\pi)^d} .
\ee
In particular, the (spectral) dimension $d$ of $\cM$ can be 
extracted the from the asymptotic density of the eigenvalues of $\Delta_g$.
However, although the spectrum of $\Delta_g$ contains a lot of information on the geometry, it 
does not quite determine it uniquely, 
and there are inequivalent but isospectral manifolds\footnote{One way to close this gap is to consider
spectral triples associated to Dirac operators \cite{Connes:1996gi}. 
In the matrix model, the geometrical information will be extracted more 
directly using the symplectic structure and the embedding
defined by the matrices $X^a$.}.

Now consider the spectral geometry of fuzzy spaces in more detail.
In the finite-dimensional case, the asymptotic density of eigenvalues strictly speaking vanishes, 
which would give the naive conclusion that fuzzy spaces (and all finite matrix geometries) have spectral dimension zero.
Of course this completely misses the point. 
The proper definition of a spectral dimension in the fuzzy case 
with Laplacian $\bDelta$ should be something like
\be
\cN_{\bDelta}(\L) \sim c_d{\rm vol}\cM \,\L^{d} \qquad\mbox{for}\,\, \L \leq \L_{\rm max}
\ee
where $\L_{\rm max}$ is the cutoff of the spectrum. This is of course a bit hard to make precise,
but the idea is clear.
Similarly, the information about the geometry of $\cM$ is encoded in the spectrum of its 
Laplacian or Dirac operator {\em below its cutoff}. Such a cutoff is in fact  
essential to obtain meaningful Seeley-de Witt coefficients in the noncommutative case, 
see  \cite{Blaschke:2010rr}.
Thus if ${\rm spec} \Box$ has a clear enough asymptotics for $\L \leq \L_{\rm max}$
and approximately coincides with ${\rm spec} \Delta_g$ for some classical manifold $(\cM,g)$ for
$\L \leq \L_{\rm max}$, then its spectral geometry is that of $\cM$.

To proceed, we need to specify a Laplacian for matrix geometries. Here the (Yang-Mills) matrix model provides 
a natural choice: For any given background configuration in the matrix model defined by
$D$ hermitian matrices $X^a$, there is a natural
matrix Laplace operator\footnote{This 
operator arises e.g. as equation of motion for the Yang-Mills matrix model.
There is also a natural matrix 
Dirac operator $\slashed{D}\Psi = \Gamma_a \left[X^a, \Psi\right]$ where $\Gamma_a$ generates the Clifford 
algebra of $SO(D)$. However we will not discuss it here.}
\be
{\bDelta} = [X^a,[X^b,.]]\d_{ab}
\label{matrix-laplacian}
\ee
which is a (formally) hermitian operator on $\cA$. We can study its spectrum and 
the distribution of eigenvalues. As we will explain below,
this Laplacian governs the fluctuations 
in the matrix model, and therefore encodes its effective geometry.
Hence if there is a classical geometry which approximates the matrix background $X^a$ up to some scale $\LNC$, 
the spectrum of its canonical (Levi-Civita) Laplacian $\Delta_g$ must approximately coincide 
with the spectrum of $\bDelta$, up to 
some possible cutoff $\L$. In particular,  there should be 
a map between classical functions and NC functions 
\be
\begin{array}{rcl}
\cI: \quad \cC_{\L}(\cM) &\to& \cA\,\,\subset \,\, \Mat(\infty,\C)\, \\
 f(x) &\mapsto& F
\end{array}
\label{quant-map} 
\ee
which approximately intertwines the Laplacians $\cI(\Delta_g f) \approx \bDelta(\cI(f))$.
Here $\cC_\L(\cM)$ denotes the space of functions on $\cM$ whose 
eigenvalues are bounded by $\L$, and $\cI$ should be injective.
The fuzzy sphere is an example where the matrix Laplacian precisely matches the 
classical Laplacian up to the cutoff. Its special symmetry is not essential here.

\section{Embedded noncommutative spaces and semi-classical limit.}

Although the idea of spectral geometry is clear and appropriate, it is very hard in practice to 
extract information on the metric from the spectrum. It would be much nicer to 
have a more direct handle on the geometry.
This can indeed be achieved, assuming that the matrix configuration
can be understood as quantization of an approximate 
classical symplectic manifold $(\cM,\theta^{\mu\nu})$.
We can then take advantage of the noncommutative structure
of the algebra encoded in the commutators, and interpret commutators as quantization of
the Poisson structure on $\cM$. 
In particular, the matrices $X^a$ will be interpreted as quantized embedding functions.
This makes the framework of matrix model much more accessible than the geometry 
of abstract NC spaces.

\paragraph{Quantization of Poisson manifolds.}

The quantization of a  Poisson (or symplectic) structure on $\cM$ is given by 
a quantization map (generalizing \eq{quant-map}) such that 
\bea
\cI: \quad \cC_{\L}(\cM) &\to& \cA\,\,\subset \,\, \Mat(\infty,\C)\, \nn\\
 f(x) &\mapsto& F \nn\\
 f g &\mapsto& F G + O(\theta), \qquad 
 \{f,g\} \mapsto -i[F,G] + O(\theta^2) .
\label{quant-map-2} 
\eea
Here $\theta$
encodes the scale of the Poisson tensor $\theta^{\mu\nu} = \{x^\mu,x^\nu\}$
in some local coordinates, and $O(\theta^2)$ stands for higher-order correction terms (which are unavoidable).
This will  allow to explicitly understand the 
geometry encoded in the matrix model background $X^a$, as explained below.
In any case, it is clear that $\theta^{\mu\nu}$ -- if it exists in nature --  
must play some dynamical physical role, which remains to be clarified.

The bottom line will be that any configurations in the matrix model which 
correspond to ``almost-commutative'' geometries can be related to this underlying
classical space using \eq{quant-map-2}. We can then talk about the {\em semi-classical limit}
of the matrix model background. This means that every matrix $F$ will be replaced by
its classical pre-image $\cI^{-1}(F) =: f$, and commutators will be replaced
by Poisson brackets. This allows to use the tools of classical differential geometry,
and provides the leading approximation of the geometry.
However one can go beyond this semi-classical limit, by defining an associative 
product on $\cC(\cM)$ via
\be
f \star g := \cI^{-1}(\cI(f) \cI(g)) .
\ee
This allows to systematically compute higher-order corrections of the 
NC case in the language of classical functions and geometry. 
The matrix model action (and any action in NC field theory) can then be considered as 
a deformed action on the underlying classical space.
One can moreover expand the star product ``formally'' in powers of $\theta$, as in deformation
quantization. This is very useful to improve the leading (semi-classical) description systematically by 
higher-order corrections in $\theta$.
In the context of noncommutative gauge theories (which arise in particular in matrix models), this leads to 
the concept of a Seiberg-Witten map \cite{Seiberg:1999vs}.


Going beyond the semi-classical limit,
the existence of a quantization map 
implies in particular a generalized Poincare-Birkhoff-Witt (PBW) property, in the sense that there should be a 
basis of $\cA$ organized e.g. as ordered polynomials in $X^\mu$ (times some cutoff function such as $e^{-x^2}$).
This means essentially that these ``independent generators'' $X^\mu$ can be ordered in some standard way. 
Hence the dimension of $\cM_\theta$ could be characterized by the minimal number of generators $X^a$ which 
generate $\cA$, which are functionally independent and satisfy such a PBW property.

\paragraph{Embedded noncommutative spaces.}

We are now ready to understand the geometric meaning of ``generic but smooth``
configurations in the matrix model.
The key is to interpret the matrices 
$X^a$ as quantization of the Cartesian embedding map of $\cM \subset \R^D$, i.e.
\be
X^a \sim x^a: \cM \hookrightarrow \R^D .
\ee
In particular, we can write
\be
\, [X^a,X^b] \sim i \{x^a,x^b\} = i \theta^{\mu\nu}\partial_\mu x^a \partial_\nu x^b
\ee
in the semi-classical limit, where $\theta^{\mu\nu}$ is the Poisson tensor in some 
local coordinates on $\cM$.
With a little more effort  \cite{Steinacker:2008ri,Steinacker:2010rh}, one can now show that
\be
\bDelta \phi \equiv  [X^a,[X^b,\phi]]\d_{ab} \sim  -\{X^a,\{X^b,\phi\}\}\d_{ab} = - e^\sigma \Delta_{G} \phi(x) 
\label{laplace-semiclass}
\ee
for any matrix resp. function $\phi \in \cA \sim \phi(x)$. 
Here $\Delta_{G}$ is the standard Laplace operator associated to the effective metric $G_{\mu\nu}$
defined as follows
\cite{Steinacker:2008ri}
\bea  
G^{\mu\nu}(x) &:=& e^{-\sigma}\,\theta^{\mu\mu'}(x) \theta^{\nu\nu'}(x) 
 g_{\mu'\nu'}(x)  
\label{G-def-general}  \\
g_{\mu\nu}(x) &:=& \partial_\mu x^a \partial_\nu x^b \d_{ab} \,\, ,
\label{g-explicit}\\
e^{-(n-1)\sigma} &:=& \frac 1{\theta^{n}}\, |g_{\mu\nu}(x)|^{-\frac 12},
\qquad \theta^n = |\theta^{\mu\nu}|^{1/2} .
\label{sigma-rho-relation} 
\eea
All of these are tensorial objects on $\cM$, e.g.
 $g_{\mu\nu}(x)$ is the metric induced on $\cM\subset \R^{D}$ via 
pull-back of $\d_{ab}$. The normalization factor 
$e^{-\sigma}$ is determined uniquely such that
\be
\frac 1{\theta^{n}} = \sqrt{|G_{\mu\nu}|}\, e^{-\sigma} ,
\label{rho-sigma-det}
\ee 
except for $n=1$ which we exclude for simplicity. This provides the desired explicit 
description of the matrix geometry at the semi-classical level. Higher-order corrections
could be computed as an expansion in $\theta$, in the spirit of deformation quantization.
This generalizes the known results for spaces with additional symmetry such as the fuzzy sphere
to the case of generic matrix geometries.

The easiest way to see \eq{laplace-semiclass}
is to consider the action for a scalar field coupled to the matrix model background
\bea
S[\varphi] &\equiv& - {\rm Tr} [X^a,\phi][X^b,\phi] \d_{ab} 
\sim \frac{1}{(2\pi)^n}\, \int d^{2n} x\; 
\sqrt{|G_{\mu\nu}|}\,G^{\mu\nu}(x)
 \partial_{\mu} \phi \partial_{\nu} \phi \,.
\label{covariant-action-scalar}
\eea
Writing the lhs as ${\rm Tr}\phi \Box\phi$, we obtain \eq{laplace-semiclass}. Moreover, 
note that $\phi$ in this action can be viewed as additional (i.e. transversal) matrix component 
$\phi \equiv X^{D+1}$ in an extended matrix model. For the same reason,
\eq{covariant-action-scalar} is precisely the action which governs e.g. nonabelian scalar fields
in the original matrix model, which arise as fluctuations of the {\em transversal} 
matrices on stacks of such backgrounds $X^a\otimes \one_n$, cf. \eq{coinciding-branes}.
This implies that these nonabelian scalar fields are governed by the effective metric $G_{\mu\nu}$.
Similarly, one can show that all fields which arise in the matrix model as fluctuations 
of the matrices around such a background (i.e. scalar fields, gauge fields and fermions) are governed by $G_{\mu\nu}$,
possibly up to a conformal factor $\sim e^\sigma$.
This means  that $G_{\mu\nu}$ is the effective gravitational metric.

We note the following observations:

\begin{itemize}
 \item 
Assume that $\dim \cM = 4$. Then $G_{\mu\nu} = g_{\mu\nu}$ if and only if the symplectic form
\be
\omega = \frac 12 \theta^{-1}_{\mu\nu} dx^\mu dx^\nu
\ee
is self-dual or anti-selfdual \cite{Steinacker:2010rh}. 

\item
There is a natural tensor
\be
\cJ^{\eta}_\gamma = e^{-\sigma/2}\, \theta^{\eta\gamma'} g_{\gamma' \gamma}
 = - e^{\sigma/2}\,  G^{\eta \gamma'} \theta^{-1}_{\gamma' \gamma} .
\label{J-tensor}
\ee
Then the effective metric can be written as
\be
G^{\mu\nu} 
= \cJ^{\mu}_\rho\, \cJ^{\nu}_{\rho'}\, g^{\rho\rho'}
= - (\cJ^2)^{\mu}_\rho\, g^{\rho\nu}.
\ee
In particular, $\cJ$ defines an almost-complex structure if and only if $G_{\mu\nu} = g_{\mu\nu}$,
hence for  (anti-)selfdual $\omega$. In that case, 
$(\cM,\tilde g,\omega)$ defines an almost-K\"ahler structure on $\cM$ where
\be
\tilde g_{\mu\nu} := e^{-\sigma/2}\,  g_{\mu\nu} .
\label{almost-K}
\ee 
\item
The matrix model is invariant under 
gauge transformations $X^a \to {X^a}' = U^{-1} X^a U$, which semi-classically 
correspond to symplectomorphisms $\Psi_U$ on $(\cM,\omega)$.
This can be viewed in terms of modified embeddings ${x^a}' = x^a \circ \Psi_U: \,\, \cM \to \R^D$
with equivalent geometry.
\item
Matrix expressions such as $[X^a,X^b] \sim i \theta^{\mu\nu}\del_\mu x^a \del_\nu x^b$ 
should be viewed as (quantizations of) tensor fields on $\cM\subset \R^D$,
written in terms of Cartesian coordinates $a,b$ of the ambient space $\R^D$.
Note that they are always tangential, because $\del_\nu x^b \in T_p\cM$.
Using appropriate projectors on the tangential resp. normal bundles of $\cM$, this can be 
used to derive matrix expressions which encode e.g. the intrinsic curvature of $\cM$,
cf. \cite{Blaschke:2010rg,Arnlind:2010kw}. This is important for gravity.

\end{itemize}

\section{Realization of certain generic 4D geometries in matrix models}
\label{sec:generic-4D}

In this section, we want to show how a large class of generic 4-dimensional geometries can be realized as NC
branes in matrix models with $D=10$. This should eliminate any lingering doubts about the 
geometrical scope of the matrix model approach to gravity.
One way to see this is as follows:
\begin{enumerate}
 \item 
Consider some ''reasonable'' generic geometry $(\cM^4,g_{\mu\nu})$ with nice properties, as explained below.
\item
Choose an embedding $\cM \hookrightarrow \R^D$. This is in general not unique, and 
requires that $D$ is sufficiently large. Using classical embedding theorems \cite{clarke}, 
$D=10$ should be enough to embed generic physically relevant 4-dimensional geometries
(at least locally).

\item
Equip $\cM$ with an (anti-)selfdual closed 2-form $\omega$. Notice that this means
$d\omega = d\star_g \omega = 0$, i.e. $\omega$ is a special solution of the free Maxwell equations on $\cM$.
Such a solution generically exists for mild assumptions on $\cM$, for example by solving the 
corresponding boundary value problem with 
$\omega$ being (anti-)selfdual on the boundary or asymptotically\footnote{However, it may happen that 
$\omega$ vanishes at certain locations, cf. \cite{Blaschke:2010ye}. 
This might be cured through compact extra dimensions.}.
The requirements in step 1) should ensure that this is possible.
For asymptotically flat spaces, $\omega$ should be asymptotically constant in order to ensure
that the dilaton $e^{-\sigma}$ is asymptotically constant.
In the case of compact extra dimensions $\cM^4 \times K$, this requirement may be relaxed.

As explained above, it follows that $(\tilde g,\omega)$ \eq{almost-K} is almost-K\"ahler. 
Under mild assumptions, one can then show 
\cite{uribe} that there exists a quantization \eq{quant-map-2} of the symplectic space $(\cM,\omega)$ in terms of 
operators on a Hilbert space\footnote{The use of the almost-K\"ahler structure 
may only be technical and should actually not be necessary.}.

In particular, we can define $X^a := \cI(x^a)\in \cA$ to be the matrix obtained as quantization of $x^a$,
so that
\be
X^a \sim x^a: \cM \to \R^D .
\ee
The effective metric on $\cM$ is therefore $G_{\mu\nu}$ as explained above.
\item
Since $\omega$ is (anti-)selfdual it follows that $G = g$, and we have  indeed obtained a quantization of $(\cM,g)$
in terms of a matrix geometry. In particular, the matrix Laplacian $\bDelta$ will approximate $\Delta_g$ for low enough
eigenvalues, and fluctuations of the matrix model around this background
describe fields propagating on this effective geometry.

\end{enumerate}

\section{Deformations of embedded NC spaces}

%
%

Assume that $X^a \sim x^a:\,\cM \hookrightarrow \R^D$ describes some quantized embedded space as before.
The important point which justifies the significance of this class of configurations
is that {\em it is preserved by small deformations}. 
Indeed, consider a small deformation $\tilde X^a =  X^a +A^a$ by generic matrices $A^a \in \cA$.
By assumption, there is a local neighborhood 
for any point $p\in \cM$ where we can separate the matrices $X^a$ into 
independent coordinates and embedding functions,
\be
X^a = (X^\mu,\phi^i(X^\mu)) 
\label{X-splitting}
\ee
such that the $X^\mu$ generate the full\footnote{In topologically non-trivial situations 
they will individually generate only ``almost`` the full $\cA$, and $\cA$
is recovered by combining various such local descriptions. This will become more clear in the example of $S^2_N$.} 
matrix algebra $\cA$.
Therefore we can write in particular  $A^a = A^a(X^\mu)$, and assume that it is smooth
(otherwise the deformation will be suppressed by the action). 
We can now consider $\tilde X^\mu = X^\mu+A^\mu \sim \tilde x^\mu(x^\nu)$ as new coordinates with modified 
Poisson structure $[\tilde X^\mu,\tilde X^\nu] \sim i \{\tilde x^\mu,\tilde x^\nu\}$, and 
 $\tilde \phi^i = \phi^i + A^i \sim \tilde \phi^i(\tilde x^\mu)$ as modified embedding 
of $\tilde \cM \hookrightarrow \R^D$. Therefore $\tilde X^a$ describes again a quantized embedded space.
This property should also ensure that embedded NC spaces play a dominant role in the path integral \eq{pathintegral}.

If we do not want to assume the existence of a quantized embedded space, things are more difficult.
The existence of a PBW property for a subset $X^\mu$ of the matrices might be a substitute, 
so that general functions $\phi(X^\mu)$ can be expanded in some basis of ''ordered`` functions of $X^\mu$,
in particular $[X^\mu,X^\nu] = i \theta^{\mu\nu}(X^\rho)$. This 
should essentially imply the existence of
some sort of quantization map  \eq{quant-map-2}.
However this seems  not very ''intrinsic``. 

One particularly interesting point is the notion of dimension, which should be the number of
independent generators in \eq{X-splitting}, or the rank of $[X^a,X^b]$. 
Semi-classically, this dimension can be extracted purely algebraically from 
$J^a_b:= -i[X^a,X^c] \d_{cb}$, which semi-classically reduces to the tensor field 
$e^{\sigma/2}\cJ^\mu_\nu$ \eq{J-tensor}. Therefore 
 it satisfies a characteristic equation of order $\dim \cM$ \cite{Steinacker:2010rh}. 
However it is not clear if this still holds e.g. for higher-order corrections in $\theta$.
If so, this would provide a very useful intrinsic characterization of matrix geometry.

To obtain an intuition and to understand the meaning of ''local description'', 
consider the example of the fuzzy sphere. 
For example, we can solve for $X^3 = \pm \sqrt{1 - (X^1)^2 - (X^2)^2}$, and use 
$X^1, X^2$ as local coordinate near the north pole $X^3 = +1$ resp. the south pole $X^3 = -1$.
Each branch of the solution makes sense provided some restriction on the spectrum of 
$X^3$ is imposed, and in general ``locality`` might be phrased as a condition on the spectrum
of some coordinate(s).
Then the $X^1,X^2$ ''locally'' generate the full matrix algebra $\cA$, and satisfy a PBW property.

%


The existence of a splitting \eq{X-splitting} can be exploited further using 
the  $ISO(D)$ symmetry of Yang-Mills matrix models. In the semi-classical picture,
one can thus assume for any given point $p \in \cM$ that $\partial_\mu \phi^i = 0$, i.e. the tangent space
is spanned by the first $d$ coordinates in $\R^D$. Moreover, $p$ can be moved to the origin using the 
$D$-dimensional translations.
Then the matrix geometry looks locally exactly 
like $\R^d_\theta$, which is deformed geometrically by non-trivial $\phi^i(X^\mu)$ and a non-trivial
commutator $[X^\mu,X^\nu] = i (\bar \theta^{\mu\nu} + \d \theta^{\mu\nu}(X^\a))$. 
These $X^\mu$ define ''local embedding coordinates``,
which are analogous to Riemannian normal coordinates.
Hence any deformation of 
$\R^d_\theta$ gives a matrix geometry as considered here, and vice versa any matrix geometry which is 
in some sense locally smooth should have such a local description. This completes the (heuristic) justification 
of our treatment of matrix geometry.

Finally, we are free in principle to use any other noncommutative ''local coordinates``,
i.e. $Y^\mu(X^\nu)$, and write the resulting action in terms of $Y^\mu$.
In the infinite-dimensional case, one can in particular try to choose\footnote{this is clearly related to 
rigidity theorems for the Heisenberg algebra.} the analog of local Darboux coordinates,
defined as $[Y^\mu,Y^\nu] = i \bar\theta^{\mu\nu}$ for constant $\bar\theta^{\mu\nu}$.
This amounts to the Moyal-Weyl quantum plane. However, the action then takes a 
highly non-trivial non-polynomial form\footnote{I would like to thank Alexander Schenkel for discussions on this
point.}.

\section{Further aspects and generalizations}

Although we focused so far on matrix geometries which are quantizations of classical 
symplectic manifolds, 
it should be stressed again that matrix models are much richer and accommodate 
structures such as multiple branes,
intersecting branes, manifolds suspended between branes, etc.  

Recall that the algebra generated by (finite-dimensional) 
hermtitian matrices $X^a$ is always a product of simple matrix algebras,
i.e. it decomposes into diagonal blocks.
One particularly simple and important case is that of coinciding branes. 
Suppose that $X^a \in \cL(\cH)$ is some matrix realization resp. quantization of 
$x^a: \cM \hookrightarrow \R^D$ as discussed above. 
Then the following configuration
\be
Y^a = X^a \otimes \one_n = \left(\begin{array}{cccc}
                           X^a & 0 & 0 & 0 \\
                           0 & X^a & 0 & 0 \\
                           0 & 0 & \ddots & 0 \\
                           0 & 0 & 0 & X^a 
                           \end{array}\right)
\label{coinciding-branes}
\ee
should be interpreted as $n$ coinciding branes. 
This is instructive because the underlying algebra
$\cA \otimes \Mat(n,\C)$ can be interpreted in
two apparently different but nonetheless equivalent ways: 1) as $su(n)$ valued functions on $\cM$
or 2) describing a higher-dimensional space $\cM \times K$, where $\Mat(n,\C)$ is 
interpreted as quantization of some compact symplectic space $K$.
Which of these two interpretations is physically correct depends on the actual 
matrix configuration, generalizing \eq{coinciding-branes}. Such extra dimensions provide a natural way of adding more 
structure to the effective physics such as physically relevant gauge groups etc., 
cf.  \cite{Aschieri:2006uw}. 

Another variation of this idea allows to describe continuous superpositions of such branes, 
which involves a classical direction.
For example, consider
$Y^a = (X^\mu \otimes \one, 1\otimes C)$ where $C$ is a selfadjoint operator with continuous 
spectrum, such as a quantum mechanical position operator. 
This would describe the geometry of $\cM \times \R$. However, in this case 
the matrix $X^i = \one\otimes C$ commutes with the $X^\mu \otimes \one$, and
$\cM \times \R$ is really a foliation with symplectic leaves 
$\{\cM \times \{c\}; \,\, c \in \R\}$ and an extra classical direction.
In that case, the effective metric $G^{\mu\nu}$ \eq{G-def-general} in the matrix model is degenerate along
the classical direction, so that fluctuations propagate only along the 
symplectic leaves. This is a very important property of the matrix model. It means that 
its effective geometry is necessarily non-commutative, justifying our focus on quantized
symplectic spaces. This is also an important difference to the standard string theory picture where
D-branes with a B-field do couple to the bulk physics. In our matrix model backgrounds, the bulk is 
essentially decoupled.

It is interesting to compare the present picture of embedded NC branes with other types of solutions of the IKKT model,
such as the compactification on noncommutative tori in \cite{Connes:1997cr}.
The latter are configurations of the type $U^{-1} X^i U =  X^i + R^i$, which allow to obtain 
10-dimensional compactified solutions such as $\R^4 \times T^6$. These solutions 
do not belong to the class of embedded NC branes considered here. They are ''space-filling`` branes,
whose tori are infinite-dimensional algebras which in some sense contains also a ''winding`` sector. 
They are not stable under small deformations (e.g. it makes an important difference whether $\theta$ is rational or irrational).
In contrast, embedded NC spaces as considered here with extra dimensions such as \eq{fuzzy-torus}
can be at most 8-dimensional. They contain no winding modes and are stable under deformations. Therefore we
are considering a different sector of the matrix model, which is better behaved in many ways.

Last but not least, it should be emphasized that the geometry and therefore gravity 
described by $X^a$ is not fixed but determined dynamically in these matrix models,  
depending notably on the presence of matter. 
The quantization of the IKKT model in terms of an integral over all 
(bosonic and fermionic) matrices \eq{pathintegral}
can be expected to be well-defined, because of maximal supersymmetry. Therefore this and related models
provide excellent candidates for a quantum theory of gravity coupled to matter.

\subsection{Acknowledgments}

I would like to thank the organizers of the 2010 Corfu Summer 
Institute of elementary particle
physics for providing a pleasant venue for stimulating and interesting 
discussions. I also want to thank Paolo Aschieri, who persistently 
confronted me with questions about basic aspects of the geometry in matrix models, 
which motivated this article. I also thank Clifford Taubes for pointing out limitations of the 
class of geometries under consideration, and   
Paul Schreivogel for useful discussions 
on the fuzzy torus.
This work was supported by the FWF project  P21610.

\end{document}